\documentclass[a4paper]{article}
\usepackage[utf8]{inputenc}
\usepackage{graphicx}
\usepackage{onecolceurws}

\usepackage{lean_unicode}
\usepackage{amsmath}
\usepackage{amsfonts}
\usepackage[autosize]{dot2texi}
\usepackage{tikz}
\usepackage{todonotes}
\usepackage{xspace}
\usepackage{xpatch}
\usepackage{cleveref}
\usepackage{hyperref}
\usetikzlibrary{shapes,arrows}
\usepackage[frozencache,cachedir=minted-cache]{minted}
\usepackage{csquotes}
\usepackage{enumitem}
\usepackage{orcidlink}
\setlist{nosep}
\setminted{
  fontsize=\footnotesize,
  frame=single,
  breaklines
}
\makeatletter
\newcommand{\currentfontsize}{\fontsize{\f@size}{\f@baselineskip}\selectfont}
\makeatother
\setmintedinline{fontsize=\currentfontsize}
\definecolor{bg}{rgb}{0.95,0.95,0.95}
\newmintinline[lean]{theorem.py:LeanLexer -x}{bgcolor=bg}
\newminted[leancode]{theorem.py:LeanLexer -x}{}

\fvset{listparameters=\setlength{\topsep}{0pt}\setlength{\parsep}{0pt}\setlength{\partopsep}{0pt}}
\AtBeginEnvironment{minted}{\vspace{4pt}}

\newlength{\mintedfboxsep}
\makeatletter
\newlength{\fboxrsep}
\newlength{\fboxlsep}
\newlength{\fboxtsep}
\newlength{\fboxbsep}
\xpatchcmd{\minted@inputpyg@inline}{%
  \colorbox%
}{%
  \long\def\color@b@x##1##2##3%
  {\leavevmode
    \setbox\z@\hbox{\kern\fboxlsep{\set@color##3\strut}\kern\fboxrsep}%
    \dimen@\ht\z@\advance\dimen@\fboxtsep\ht\z@\dimen@
    \dimen@\dp\z@\advance\dimen@\fboxbsep\dp\z@\dimen@
    {##1{##2\color@block{\wd\z@}{\ht\z@}{\dp\z@}\box\z@}}}%
  \colorbox%
}{\typeout{Success}}{\typeout{Failure}}
\makeatother

\title{Scalar actions in Lean's \mathlib}

\author{
Eric Wieser\,\orcidlink{0000-0003-0412-4978} \\ efw27@cam.ac.uk
}

\institution{Cambridge University Engineering Department}

\newcommand{\mathlib}{\textsf{mathlib}\xspace}

\begin{document}
\maketitle

\begin{abstract}
Scalar actions are ubiquitous in mathematics, and therefore it is valuable to be able to write them succinctly when formalizing.
In this paper we explore how Lean 3's typeclasses are used by \mathlib for scalar actions with examples, illustrate some of the problems which come up when using them such as compatibility of actions and non-definitionally-equal diamonds, and note how these problems can be solved.
We outline where more work is needed in \mathlib in this area.
\end{abstract}
\vskip 32pt

\section{Introduction}
In this paper, we explore some of the design decisions made in \mathlib \cite{The_mathlib_Community_2020}, a mathematical library for the Lean 3 theorem prover \cite{demoura}.
In particular, we look at the scalar action operator \lean{•} as a case-study in how typeclasses are used.
In \cref{sec:has_scalar} we introduce the typeclass which provides this notation, and some of the hierarchy of stronger typeclasses that surround it.
In \cref{sec:elementary} we outline some basic actions, and in \cref{sec:derived} show how these are used to build more complex classes.
In \cref{sec:algebra} we show how some of the extra compatibility typeclasses associated with these actions can be repurposed to work with various restricted versions of unital associative algebras, referencing two recent contributions to \mathlib that exploited this approach.
In \cref{sec:diamonds} we provide a brief example of how typeclass diamonds can arise, and why they matter.
We conclude in \cref{sec:future} by discussing further work needed on \mathlib, primarily regarding scalar \emph{right}-actions.

\subsection{Typeclasses}
\label{sec:Typeclasses}
A central language feature used by \mathlib in expressing algebraic structure is that of typeclasses \cite[section~4]{The_mathlib_Community_2020}, which are used to equip types with canonical operators and properties of those operators.
A simple example is the typeclass \lean{semigroup M}, which equips the type \lean{M} with the operator \lean{*} and an associativity axiom \lean{mul_assoc}.
Working with typeclasses breaks down into three parts; declaring them with the \lean{class} keyword, providing them with the \lean{instance} keyword, and consuming them with \lean{[ ]} around a typeclass name.

The \lean{semigroup} typeclass from \mathlib is declared roughly as follows, which reads \enquote{a type \lean{G} has a multiplication if it has a binary operator called \lean{mul}} and \enquote{a type \lean{G} is a semigroup if it has a multiplication and that multiplication is associative}.
Note that the use of \lean{extends} additionally tells Lean how to obtain an instance of \lean{has_mul G} if it has \lean{semigroup G}.
\begin{leancode}
class has_mul (G : Type*) :=
(mul : G → G → G)

infix * := has_mul.mul
\end{leancode}
\begin{minipage}{\textwidth}  
\vspace{-4pt}
\begin{leancode}
class semigroup (G : Type*) extends has_mul G :=
(mul_assoc : ∀ a b c : G, a * b * c = a * (b * c))
\end{leancode}
\vspace{4pt}
\end{minipage}

With this typeclass in place, we can write a theorem that applies to any semigroup by writing \lean{[semigroup G]} in our argument list, as we do in \lean{mul_assoc₂} below.
This tells Lean that whenever the \lean{mul_assoc₂} lemma is used on a type \lean{G}, it should perform a typeclass search for a term of type \lean{semigroup G}.
In turn, it means that inside \lean{mul_assoc₂} we have access to the \lean{mul_assoc} axiom of semigroups on \lean{G}.
\begin{leancode}
lemma mul_assoc₂ {G : Type*} [semigroup G] (a b c d : G) : a * (b * (c * d)) = ((a * b) * c) * d :=
by rw [mul_assoc, mul_assoc]
\end{leancode}

The final piece of the puzzle is how to inform the typeclass search that \lean{semigroup G} is available for a particular type \lean{G}.
To demonstrate this, we define a structure \lean{opposite α} that wraps a single element of an arbitrary type \lean{α}.
Using the \lean{instance} keyword, we then equip it with a reversed multiplication structure, and express \enquote{For any type \lean{α} such that \lean{α} is itself a semigroup, \lean{opposite α} is also a semigroup}.
This method of \enquote{chaining} instances is central to the power of typeclasses, and is used extensively by \mathlib in situations like equipping a product of groups with a group structure, or polynomials over a ring with a ring structure.
\begin{leancode}
structure opposite (α : Type*) := (x : α)

instance (T : Type*) [semigroup T] : semigroup (opposite T) :=
{ mul := λ a b, ⟨b.x * a.x⟩,
  mul_assoc := λ a b c, congr_arg opposite.mk (mul_assoc c.x b.x a.x).symm }
\end{leancode}

\subsection{The \texorpdfstring{\lean{has_scalar}}{has\_scalar} typeclass}
\label{sec:has_scalar}
The typeclass we are most interested in this paper is \lean{has_scalar M α}, which equips a type \lean{α} with an action by elements of \lean{M} denoted \lean{m • a}.
In practice, this is almost always used for group actions, which are actions that satisfies the additional fields in \lean{mul_action M α}:
\begin{leancode}
class has_scalar (M : Type*) (α : Type*) := (smul : M → α → α)

infixr ` • `:73 := has_scalar.smul

class mul_action (M : Type*) (α : Type*) [monoid M] extends has_scalar M α :=
(one_smul : ∀ a : α, (1 : M) • a = a)
(mul_smul : ∀ (x y : M) (a : α), (x * y) • a = x • y • a)
\end{leancode}
Note here that because we use \lean{[monoid M]} instead of 
\lean{extends monoid M}, we are stating that \lean{mul_action M α} requires \lean{M} to already be equipped with a monoid structure, rather than allowing \lean{mul_action M α} to itself provide that structure.

\mathlib extends these two typeclasses with a variety of additional axioms (i.e., fields holding proofs) for when \lean{M} and \lean{α} are themselves equipped with extra structures, such as distributivity over addition and actions by zero.
The left hand side of \cref{fig:hierarchy} shows the majority of these typeclasses, while details of their fields can be found either in \cite[section~5.1]{The_mathlib_Community_2020} or in the \mathlib docs.

\begin{figure}[b]
\begin{minipage}[b]{0.6\textwidth}
\centering
{
\begin{dot2tex}[dot,options=--cache]
digraph G {

node [shape=none];
ranksep="0.1"
nodesep="0.1"

subgraph cluster_0 {
    peripheries = 0
    labelloc="b";
    
    has_scalar [texlbl="\scriptsize\lean{has_scalar}"];
    mul_action [texlbl="\scriptsize\lean{mul_action}"];
    smul_with_zero [texlbl="\scriptsize\lean{smul_with_zero}"];
    mul_action_with_zero [texlbl="\scriptsize\lean{mul_action_with_zero}"];
    distrib_mul_action [texlbl="\scriptsize\lean{distrib_mul_action}"];
    module [texlbl="\scriptsize\lean{module}"];
    algebra [texlbl="\scriptsize\lean{algebra}"];
    mul_semiring_action [texlbl="\scriptsize\lean{mul_semiring_action}"];
    
    smul_with_zero -> has_scalar;
    mul_action -> has_scalar;
    mul_action_with_zero -> mul_action;
    mul_action_with_zero -> smul_with_zero;
    distrib_mul_action -> mul_action;
    module -> distrib_mul_action;
    module -> mul_action_with_zero;
    algebra -> module;
    mul_semiring_action -> distrib_mul_action;

};

subgraph cluster_1 {
    peripheries = 0
    labelloc="b";
    has_mul [texlbl="\scriptsize\lean{has_mul}"];
    mul_zero_class [texlbl="\scriptsize\lean{mul_zero_class}"];
    monoid [texlbl="\scriptsize\lean{monoid}"];
    monoid_with_zero [texlbl="\scriptsize\lean{monoid_with_zero}"];
    semiring [texlbl="\scriptsize\lean{semiring}"];
    comm_semiring [texlbl="\scriptsize\lean{comm_semiring}"];

    mul_zero_class -> has_mul;
    monoid -> has_mul;
    monoid_with_zero -> mul_zero_class;
    monoid_with_zero -> monoid;
    semiring -> monoid_with_zero;
    comm_semiring -> semiring;
    
    has_mul -> has_scalar [constraint=false, color="gray"];
    mul_zero_class -> smul_with_zero [constraint=false, color="gray"];
    monoid -> mul_action [constraint=false, color="gray"];
    monoid_with_zero -> mul_action_with_zero [constraint=false, color="gray"];
    semiring -> module [constraint=false, color="gray"];
    comm_semiring -> algebra [constraint=false, color="gray"];
}

}
\end{dot2tex}
}
\caption{Hierarchy of scalar action typeclasses}
\label{fig:hierarchy}
\medskip
\small
Arrows indicate implications.
Grey arrows indicate implied left-multiplication actions.
\end{minipage}%
\begin{minipage}[b]{0.4\textwidth}
\centering
{\small
\begin{dot2tex}[dot,options=--cache]
digraph G {

node [shape=none];
ranksep="0.1"
nodesep="0.1"
    
    addiAB [texlbl="\scriptsize\lean{add_comm_monoid (ι → A →+ B)}"];
    addAB [texlbl="\scriptsize\lean{add_comm_monoid (A →+ B)}"];
    addB [texlbl="\scriptsize\lean{add_comm_monoid B}"];
    moduleiAB [texlbl="\scriptsize\lean{module ℕ (ι → A →+ B)}"];
    moduleAB [texlbl="\scriptsize\lean{module ℕ (A →+ B)}"];
    moduleB [texlbl="\scriptsize\lean{module ℕ B}"];
    
    addB -> moduleB -> moduleAB -> moduleiAB [color=red]
    addB -> addAB -> moduleAB -> moduleiAB [color=green]
    addB -> addAB -> addiAB -> moduleiAB [color=blue]
}
\end{dot2tex}
}
\caption{Diamonds in typeclass search}
\label{fig:diamonds}
\medskip
\small
Arrows indicate implications, showing the three possible paths to resolve \lean{module ℕ (A →+ B)}.
\end{minipage}
\end{figure}

\section{Elementary actions}
\label{sec:elementary}
Scalar actions can be roughly divided into two types: elementary actions which are intrinsic to a particular family of types, and derived actions which operate elementwise on \enquote{bigger} types built out of smaller types.
We will start by giving some examples of the former.

\subsection{Left multiplication}
\label{sec:mul_to_smul}

One of the simplest actions we can construct is that of left-multiplication, with \lean{a * b = a • b}, which \mathlib provides as follows.
\begin{leancode}
instance has_mul.to_has_scalar (α : Type*) [has_mul α] : has_scalar α α := { smul := (*) }
\end{leancode}
As the properties of the multiplication on \lean{α} becomes stronger, so do those of this scalar action on \lean{α}; for instance when we have \lean{monoid α} we can deduce \lean{mul_action α α}, and when we have \lean{semiring α} we can deduce \lean{module α α}.
The right-hand side of \cref{fig:hierarchy} shows these available left multiplication structures with grey arrows.

\subsection{Repeated addition and subtraction}
\label{sec:repeated_addition}

Another simple action we can construct is that of repeated addition when \lean{α} is a commutative additive monoid, an instance of \lean{module ℕ α}, which can be defined recursively for a natural number as \lean{(0 : ℕ) • x = 0} and \lean{∀ n : ℕ, (n + 1) • x = n • x + x}.
A similar approach can be used to define a \lean{module ℤ α} instance when \lean{α} additionally forms an additive group.
These are respectively promoted to \lean{algebra ℕ α} and \lean{algebra ℤ α} structures when \lean{α} forms a \lean{semiring} or \lean{ring}.

\section{Derived actions}
\label{sec:derived}

A typical example of a module action might be that of a scalar $\mathbb{R}$ on the vector space $\mathbb{R}^3$ (\lean{fin 3 → ℝ}), which multiplies each component separately.
After making the obvious generalization to an arbitrary type and index set, the easy way to write this down would be as follows, where again we can provide a stronger \lean{module α (ι → α)} if we know \lean{α} forms a \lean{semiring}.
\begin{leancode}
instance function.has_scalar (ι α : Type*) [has_mul α] : has_scalar α (ι → α) :=
{ smul := λ r v, (λ i, r * v i)}
\end{leancode}
This definition is perfectly fine for the action we wanted, but we can still generalize it much more.
Consider now the action on matrices \lean{ι₁ → ι₂ → R} by their coefficients \lean{R}.
We would like to show \lean{has_scalar R (ι₁ → ι₂ → R)}, but that doesn't match the \lean{function.has_scalar} instance we just defined.
While we could obviously define this operation trivially just as we did there, we would have to do so again if working with a vector of matrices or similar.

A better approach here is to exploit the chaining that occurs during typeclass search, and define our action as:
\begin{leancode}
instance function.has_scalar' (ι M α : Type*) [has_scalar M α] : has_scalar M (ι → α) :=
{ smul := λ r v, (λ i, r • v i) }
\end{leancode}
This instance is strictly more general---typeclass search will recover our original \lean{has_scalar α (ι → α)} instance by setting \lean{M = α} and finding \lean{has_scalar α α} from \lean{has_mul.to_has_scalar}, but can also find the \lean{has_scalar R (ι₁ → ι₂ → R)} we wanted by setting \lean{M = R} and  \lean{α = (ι₂ → R)}, and finding \lean{has_scalar R (ι₂ → α)} by recursive application of this instance. This approach is used extensively throughout \mathlib, for actions on
\begin{enumerate}
\item sets and products defined in terms of actions on their elements
\item polynomials defined in terms of actions on their coefficients
\item bundled homomorphisms defined in terms of actions on their codomain
\end{enumerate}
Most of these actions propagate their axioms; for instance when we used \lean{[has_scalar M α]} to define \lean{has_scalar M (ι → α)}, we can show that if we additionally have \lean{[module M α]} to define \lean{module M (ι → α)}.

\subsection{More complex derived actions}

In \cref{sec:derived}, the action we describe contains no proof obligations---we did not need to know any properties of \lean{[has_scalar M α]} to define \lean{has_scalar M (ι → α)}.
Sometimes, the typeclasses in \cref{fig:hierarchy} are enough to resolve this---for instance, while we can't conclude \lean{has_scalar R (M →+ N)} from \lean{[has_scalar R N]} as we don't know enough about this action to know if additive maps remain additive, we can conclude \lean{distrib_mul_action R (M →+ N)} from \lean{[distrib_mul_action R N]}.

Once we start working with types that themselves ingrain a preferred action though, we need some additional tools.
For instance, the closely related types for \lean{R}-linear maps \lean{M →ₗ[R] N} and \lean{R}-submodules \lean{submodule R N} ingrain a preferred \lean{R}-action.
We can start by attempting to a general action by an arbitrary type \lean{α}.
If we do this we find ourselves left with two proof obligations, indicated by the \lean{show ..., from} syntax.
\begin{leancode}
instance {α R M N : Type*}
  [semiring R] [add_comm_monoid M] [add_comm_monoid N] [has_scalar α N] [module R M] [module R N] :
  has_scalar α (M →ₗ[R] N) :=
{ smul := λ a f, { to_fun := λ m, a • f m,
                   map_add' := λ m₁ m₂, (congr_arg _ $ f.map_add _ _).trans $
                     show a • (f m₁ + f m₂) = a • f m₁ + a • f m₂, from sorry,
                   map_smul' := λ r m, (congr_arg _ $ f.map_smul _ _).trans $
                     show a • r • f m = r • a • f m, from sorry } }
\end{leancode}
The goal in \lean{map_add'} tells us we need to strengthen \lean{[has_scalar α N]} to \lean{[monoid α]} \lean{[distrib_mul_action α N]}.

The goal in \lean{map_smul'} is more troublesome.
The easy way out is to replace \lean{α} with a commutative \lean{R} so our statement becomes
\begin{leancode}
instance {α M N : Type*} [comm_semiring R] [add_comm_monoid M] [add_comm_monoid N] [module R M] [module R N] :
  has_scalar R (M →ₗ[R] N) :=
\end{leancode}
and the \lean{sorry} can be closed with \lean{a • r • f m = (a * r) • f m = (r * a) • f m = r • a • f m} which follows from the axioms of \lean{mul_action} and commutativity of \lean{R}.
Another approach would be to require \lean{R} to be an \lean{α}-algebra \lean{[algebra α R]}, and that the \lean{α}-action on \lean{R} and \lean{N} is compatible with the \lean{R}-action on \lean{N}.

\label{sec:is_scalar_tower}
To best solve this problem, \mathlib provides two additional typeclasses about scalar actions.
The first expresses the compatibility condition we would need to use \lean{[algebra α R]} as mentioned above, as
\begin{leancode}
class is_scalar_tower (M N α : Type*) [has_scalar M N] [has_scalar N α] [has_scalar M α] : Prop :=
(smul_assoc : ∀ (x : M) (y : N) (z : α), (x • y) • z = x • (y • z))
\end{leancode}
The name alludes to towers of algebras, which is described in more detail in \cite[Section~3.2]{baanen2021formalization}.
Our particular problem can be solved more directly with the second typeclass, \lean{[smul_comm_class α R N]}, which expresses exactly the condition we require:
\begin{leancode}
class smul_comm_class (M N α : Type*) [has_scalar M α] [has_scalar N α] : Prop :=
(smul_comm : ∀ (m : M) (n : N) (a : α), m • n • a = n • m • a)
\end{leancode}
After this typeclass was introduced in \cite{pr-4770}, the author contributed and drove the review of a large number of instances of it, most notably those for polynomials, product types, and the repeated addition actions in \cref{sec:repeated_addition}.

\section{Algebras and not-quite-algebras}
\label{sec:algebra}

The \mathlib \lean{algebra R A} describes an associative unital \lean{R}-algebra over \lean{A} given a \lean{comm_semiring R} and \lean{semiring A}.
The definition is roughly
\begin{leancode}
class algebra (R A : Type*) [comm_semiring R] [semiring A] extends has_scalar R A :=
(algebra_map : R →+* A)
(commutes : ∀ r x, algebra_map r * x = x * algebra_map r)
(smul_def : ∀ r x, r • x = algebra_map r * x)
\end{leancode}
which states that there is a canonical ring homomorphism from \lean{R} to \lean{A} which agrees with \lean{•} and sends \lean{R} to the center of \lean{A}.
This parameterization of the axioms is difficult to generalize to A being nonunital and non-associative ring.
However, \mathlib also provides this definition to construct an algebra from an alternate set of axioms:
\begin{leancode}
def algebra.of_module (R A : Type*) [comm_semiring R] [semiring A] [module R A]
  (h₁ : ∀ (r : R) (x y : A), (r • x) * y = r • (x * y))
  (h₂ : ∀ (r : R) (x y : A), x * (r • y) = r • (x * y)) : algebra R A := sorry
\end{leancode}
If we look carefully, we note that \lean{h₁} and \lean{h₂} closely resemble \lean{smul_assoc} and \lean{smul_comm} from \cref{sec:is_scalar_tower}, but with some \lean{*}s substituted for \lean{•}.
But if we look back to \cref{sec:mul_to_smul}, we remember that when \lean{x} and \lean{y} are the same type, \lean{x * y = x • y} by definition!
This means that \lean{h₁} and \lean{h₂} correspond directly with \lean{is_scalar_tower R A A} and \lean{smul_comm_class R A A}, respectively.

This is a valuable insight, because it allows us to use the follow sequences of typeclass arguments interchangeably:
\begin{leancode}
variables [comm_semiring R] [semiring A] [algebra R A]
\end{leancode}
\begin{leancode}
variables [comm_semiring R] [semiring A] [module R A] [is_scalar_tower R A A] [smul_comm_class R A A]
\end{leancode}
Knowing this, it becomes immediately obvious how to generalize various statements to non-unital algebras (which were needed in \cite{pr-7932}); we switch from from the first form to the second form, and then replace \lean{[semiring A]} with \lean{[non_unital_semiring A]}, something which was not permitted on the unexpanded version.
Another generalization this permits is one that allows putting \enquote{most of} an \lean{R'}-algebra structure on \lean{A} when \lean{R'} is only a monoid, which comes up for instance when \lean{R' = units R}.
In this case, we replace \lean{[comm_semiring R]} \lean{[semiring A]} \lean{[module R A]} with \lean{[monoid R]} \lean{[semiring A]} \lean{[distrib_mul_action R A]}.
This generalization was used when proving intermediate results needed for Sylvester's law of inertia \cite{pr-7416}.

\section{Diamonds}
\label{sec:diamonds}
Frequently, there are multiple ways for Lean to construct a typeclass.
For instance, consider the problem:
\begin{leancode}
example {ι A B} [add_comm_monoid A] [add_comm_monoid B] : module ℕ (ι → A →+ B) := by apply_instance
\end{leancode}
Depending on the order of the search, Lean could take any of the paths in \cref{fig:diamonds}.
While Lean does not care about the existance of multiple paths and will happily just pick one, for the typeclass to be useful we need it to be predictable to the user---all they see is a \lean{•} in the goal state.
This means that whenever we have a diamond, we want all the paths to produce the same \lean{•} such that the actual path taken does not matter.

There are two relevant notions of \enquote{same} here.
The first is propositional equality, which for the case in \cref{fig:diamonds} is easy to show as roughly \lean{∀ (M : Type*) [add_comm_monoid M], subsingleton (module ℕ M)}, that is that all \lean{ℕ}-module structures are equal.
This is enough to convince the user that the goal they're looking at is mathematically the one they're interested in.
The second is definitional equality, which is needed by Lean in order to allow lemmas about one path for the tree in \cref{fig:diamonds} to apply for lemmas about another path.

In older versions of \mathlib, the diamonds in \cref{fig:diamonds} resulted in paths creating instances that were propositionally equal, but not definitionally equal.
This was problematic, as lemmas about the natural \lean{ℕ}-action (blue path, \cref{fig:diamonds}) such as \lean{∑ x in s, c = s.card • c} would fail to match goals containing a derived \lean{ℕ}-action (green and red paths, \cref{fig:diamonds}).
This was fixed in \cite{pr-7084} by requiring the definition of \lean{add_comm_monoid M} to include an implementation of the \lean{ℕ}-module structure.
While mathematically it is bizarre to say \enquote{a commutative additive monoid has a zero, addition, \emph{and a scalar-multiplication by naturals}, such that \dots}, in Lean this is crucial to allow manual control of definitional equality such that the green and blue paths in \cref{fig:diamonds} can be made definitionally equal to the red path.
This is analogous to the situation described in \cite[section~4.1]{The_mathlib_Community_2020} for topologies associated with metric spaces.

\section{Future work}
\label{sec:future}

\subsection{Right actions}
The scalar action typeclass in \mathlib is intended for left-actions, which is apparent both in the definition of the \lean{mul_smul} axiom, and in the order in which the arguments appear in the notation.
However this does not mean that right actions are impossible.

The author has introduced preliminary support for right actions in \cite{pr-7630}, via the instance
\begin{leancode}
instance monoid.to_opposite_mul_action [monoid α] : mul_action (opposite α) α :=
{ smul := λ c x, x * c.unop,
  one_smul := mul_one,
  mul_smul := λ x y r, (mul_assoc _ _ _).symm }
  
lemma op_smul_eq_mul [monoid α] {a b : α} : op a • b = b * a := rfl
\end{leancode}
Here, \lean{opposite α} is a \mathlib type built similarly to the pedagogical example in \cref{sec:Typeclasses} which reverses the multiplication order. This permits us to write \lean{op a • b} as a messy spelling for a right action on \lean{b} by \lean{a}.
Similar instances were introduced for the other stronger typeclasses in \cref{fig:hierarchy}.
With these instances in place, it is possible to express an \lean{R}-\lean{S}-bimodule structure over \lean{M} as
\begin{leancode}
variables [module R M] [module (opposite S) M] [smul_comm_class R (opposite S) M]
\end{leancode}
Future work in this area could go on to define a \lean{subbimodule R S M}, and use this to define a \lean{two_sided_ideal R = subbimodule R R R} over a non-commutative ring \lean{R}.

Unfortunately, \mathlib does not have typeclasses for another common interaction of right actions with left actions.
Introducing briefly for clarity the notation \lean{a •> b} for \lean{a • b} and \lean{a <• b} for \lean{op a • b}, there is no typeclass capable of expressing \lean{(a <• b) •> c = a •> (b •> c)}.
Some examples of when this situation arises are \lean{[monoid M] (a b c : M)} (all three variables belong to the same non-commutative monoid), \lean{[monoid M] (a c : M) (S : submonoid M) (b : S)} (the second belongs to a submonoid of the monoid containing the other two), and \lean{[monoid M] (a b : M) (c : ι → M)} (the third variable is a coordinate vector).

The future of right actions in \mathlib might be improved by the eventual switch to Lean 4 \cite{lean4} (or the backport of design decisions made there), which provides a new \lean{HMul A B C} typeclass which makes the \lean{*} operator operate on fully heterogenous types.
One possible design choice would be to eliminate the \lean{•} operator entirely and use \lean{*} for both the left and right actions, which would make the expression above expressible as a hypothetical heterogenous semigroup axiom that would subsume \lean{is_scalar_tower}.

\subsection{Further diamond definitional alignment}
While \cite{pr-7084} fixes \lean{ℕ}-module diamonds (and a follow-up contribution fixes \lean{ℤ}-module diamonds), these problem still exist for \lean{ℕ}-, \lean{ℤ}-, and \lean{ℚ}-algebras.
This could likely be resolved by adding the new data fields \lean{of_nat},  \lean{of_int}, and \lean{of_rat} to \lean{semiring}, \lean{ring}, and \lean{division_ring} respectively, along with corresponding proof fields showing these satisfy suitable constraints.




\bibliographystyle{alphaurl} 
\bibliography{references}

\end{document}